# Query Expansion in Information Retrieval Systems using a Bayesian Network-Based Thesaurus

**Luis M. de Campos, Juan M. Fernández, Juan F. Huete**
Departamento de Ciencias de la Computación e I.A.
E.T.S.I. Informática, Universidad de Granada
18071 - Granada, SPAIN
(lci@decsai.ugr.es, jmfl@aliatar.ugr.es, jhg@decsai.ugr.es)

## Abstract

Information Retrieval (IR) is concerned with the identification of documents in a collection that are relevant to a given information need, usually represented as a query containing terms or keywords, which are supposed to be a good description of what the user is looking for. IR systems may improve their effectiveness (i.e., increasing the number of relevant documents retrieved) by using a process of query expansion, which automatically adds new terms to the original query posed by an user. In this paper we develop a method of query expansion based on Bayesian networks. Using a learning algorithm, we construct a Bayesian network that represents some of the relationships among the terms appearing in a given document collection; this network is then used as a thesaurus (specific for that collection). We also report the results obtained by our method on three standard test collections.

## 1 INTRODUCTION

A good definition that establishes the whole process of Information Retrieval (IR) is given by G. Salton [25]: *'IR is concerned with representation, storage, organisation, and accessing of information items'*. When all these tasks are carried out by means of a computer, we will refer to 'automatic information retrieval' and we will define an Information Retrieval System (IRS) as the software that implements these tasks in a computer. With respect to the meaning of an information item, we will only deal with documents, or in a broader sense, textual representations of any type of object, i.e. a research article, a book, a message in an electronic mail file, etc.

In this paper, we mainly focus our attention in the part of an IRS devoted to accessing to information items, i.e., the identification of documents in a collection that are relevant to a particular information need: an user interacts with the IRS by formulating a query, which is a description of his/her information need, getting, as a result, a set of documents which are intended to be the most suitable for the request. All IRSs draw conclusions about the content of a document (and therefore about the appropriateness of this document with respect to a given query) by examining some representation of that document, which consists of several document features, usually in the form of particular words and phrases which are intended as a good concise description of the content of the document.

To solve the IR problem, a great number of retrieval models, i.e., specifications about how to represent documents and queries and how they are compared, have been developed, since it was set up in the 1940s. In [1, 8], the reader can find a very complete review of most of the techniques used in the IR field. It should be noted that the IR problem is pervaded with uncertainty, most of the tasks involved in this area could be described as uncertain processes. In fact, the user's query is a vague description of his/her information need because, due to several reasons, he/she can not express in a precise way what he/she really wants. The query and document representation construction is another example, because it gives as a result incomplete characterisations, in the form of keywords or terms, of the content of both queries and documents.

For any given retrieval model, there is a great variety of techniques designed to improve the effectiveness of the retrieval. Among them, we stand out the use of thesauri and query expansion. A typical thesaurus contains a set of words and the relationships among them. The user of the IRS can use this structure to build queries or improve them (in the sense of getting new words to formulate the query), restricting the query if it retrieves too many documents, or broadening it if it gets too few ones. On the other hand, query expansion is the addition of new words to the query by the IRS, based on previously retrieved documents or on thesauri, having as an objective increasing the number of relevant documents retrieved.

The purpose of this research is to explore the possibilities of using Bayesian networks to automatically construct, for any given document collection, a thesaurus which may be used to improve the performance of an IRS by means of the query expansion technique (the thesaurus itself being a Bayesian network). More precisely, from the set of terms appearing in a document collection, we build a Bayesian network that represents some of the relationships among these terms using a learning algorithm. So, given a particular query, we instantiate the terms that compose it and propagate this information through the network. Then, by selecting the new terms whose posterior probability is high and adding them to the original query, we obtain the expanded query. In our experiments, we have used this method (on several standard test collections) as a preprocessing step for a classic system in the IR context, and one of the most used experimental IRSs: *Smart*. It was proposed by Salton [25] and developed at the Cornell University.

The paper is divided into 5 sections. Section 2 contains a short review of several basic concepts and methods used in Information Retrieval, which can help non-specialized readers to better understand the rest of the paper. In Section 3 we explain the learning algorithm used to construct the Bayesian network representing the thesaurus. Section 4 describes the experiments we have carried out with three standard test collections (*Adi, Cranfield and Medlars*), as well as the results obtained. Finally, Section 5 contains the concluding remarks and some proposals for future research.

## 2  PRELIMINARIES: INFORMATION RETRIEVAL BASICS

First, we are going to describe the operation of an IRS. Given a set of documents in their original format, the first step is to translate each document to a suitable representation for a computer to use. That translation is called indexing, and the output of this process is a list of words, known as terms or keywords, extracted from the text and considered significant. Sometimes, the terms have associated a number which may represents the frequency of occurrence of that term, or a more complex weight expressing the importance of the term. The IRS stores these representations instead of the complete documents, so for each document, a list of terms is stored. An inverted file is a structure in which each term has a list of documents where it occurs.

The user expresses his/her information needs formulating a query, using a formal query language or natural language. That query is also indexed to get a query representation and the retrieval continues with the part of the process in which the query representation is matched with the stored document representations using a search strategy. Finally, a set of document identifiers is presented to the user. In an operational system, the process stops here, but in an experimental one, the next step is to measure the effectiveness of the retrieval, i.e. IRS evaluation. That evaluation can be carried out with different methods, but the main one is that based on *recall* and *precision* estimations [21, 25]. The first one measures the ability of the IRS to present all the relevant documents (recall = number of relevant documents retrieved / number of relevant documents). The second one, precision, measures its ability to present only the relevant documents (precision = number of relevant documents retrieved / number of documents retrieved). To compare IRS performance in terms of these measures, the curve recall-precision is plotted for each query.

There are four classic retrieval models: Boolean, vector space, cluster and probabilistic models [21, 25]. As Boolean and cluster models are not directly related to our work, we shall briefly describe only vector space and probabilistic models. In the vector-space model, documents and queries are both represented as vectors of length $n$, being $n$ the number of terms in the collection, each position containing a weight. Documents and queries are compared using a similarity function, the most common one being the cosine of the angle between the two vectors. *Smart*, the IRS we are going to use as the basic method, belongs to this kind of models. The probabilistic retrieval also represents the documents and the queries as vectors containing a probabilistic weight for each term, which expresses the degree of importance of that term. These models compute the probability of relevance given a document and a query (the probability that a document satisfies a query) and are based on the 'Probability ranking principle'. This principle states that the best overall retrieval effectiveness will be achieved when documents are ranked in decreasing order of their probability of relevance [23]. Many probabilistic IR models have been developed, which mainly differ in the way in which they estimate the probability of relevance [12, 17, 21, 22].

As a brief bibliographical review, we are going to mention the main research lines related to applying Bayesian networks to IR. We will start at Croft and Turtle's studies [7, 28], where these authors have carried out one of the most important works in this field, developing a complete IR model based on two networks, the document and query networks, linked between them. Closely related to this work, Ghazfan et al. [13] give a different meaning to Croft and Turtle's network. Fung et al. [10] build a knowledge base, also composed of two types of Bayesian networks, that represents the relationships among concepts in the documents. Other interesting works on Bayesian networks and IR can be found in [2, 9, 18, 24].

There are several techniques which can be used in combination with any IR model to improve the effectiveness of the retrieval. We shall mention only thesauri and query expansion-based methods.

Thesauri are useful tools for the indexing and retrieval processes. A thesaurus may contain a set of basic terms and synonyms, antonyms, more general terms, more specific terms, etc, but usually in specific subject areas. Schütze and Pederson [26] make a good review of the different approaches to build these structures. We are going to center on automatic thesaurus construction. Mainly, these methods are based on statistical techniques, which try to find patterns of cooccurrence between terms among all the documents in the collection. This task is carried out by computing any kind of similarity function between all pairs of terms, and grouping in common classes all terms whose similarity coefficients are sufficiently large [25].

Classified as a query modification technique, query expansion is the addition of new terms to the query, based on previously retrieved documents or on thesauri, with the objective of recall improvement. Using the retrieved documents, those terms that appear in relevant documents are added to the query, and using a thesaurus, synonyms or broader terms, related terms in general, are added to query terms [15]. The basis for the cooccurrence-based thesaurus use in query expansion is the Association Hypothesis [21], that states that terms which occur in common documents tend to be about the same subject. Therefore, if a query term is good at distinguishing relevant and not relevant documents, any other term closely related to it, i.e., which cooccurs frequently with it, is also likely to be good discriminating documents and should be added to the query. Han et al. [14] present a good classification of query expansion methods.

## 3  THE THESAURUS CONSTRUCTION ALGORITHM

Given a document collection, we have built a thesaurus based on a Bayesian network. From an inverted file used as a learning file, our method learns a polytree of terms, i.e. a directed acyclic graph (dag) where there is no more than one undirected path connecting each pair of nodes. The polytree nodes represent terms of the collection in the form of binary variables. Each variable, $\alpha$, takes its values from the set $\{\alpha_0, \alpha_1\}$, where $\alpha_0$ stands for 'the term $\alpha$ is not relevant', and $\alpha_1$ represents 'the term $\alpha$ is relevant'.

There are two important reasons for learning a polytree instead of a more general Bayesian network: any document collection contains a very big number (typically thousands) of terms. As each term is associated to a node, our network shall be very large. So, the processes needed to estimate a network from empirical data (learning) can be extremely time-consuming. This task can be obviously alleviated by applying efficient learning algorithms [3, 20] which build more straightforward graphs as singly connected networks (forest, trees and polytrees), intended as approximations of more complex models due to the loss of expressiveness (since the kind of dependency/independency relationships that singly connected networks may represent is more restrictive than for general dags). The second reason is based on the methods of inference available for Bayesian networks (i.e., the propagation algorithms). Again, we deal with a potentially time-consuming process, but for singly connected networks there are also efficient and exact propagation methods which run in a time proportional to the number of nodes [19]. Among the singly connected networks, our choice was polytrees because we gain a little more accuracy. Therefore, our aim is to approximate the real dependency model of the terms in a document collection by means of a polytree, in which the most important (in)dependencies between the terms of the collection are expressed.

The algorithm we have implemented to learn the polytree is described in Figure 1. Our algorithm is quite similar to two algorithms for learning polytrees, the PA algorithm [3] and Rebane and Pearl's (RP) [20] (the two algorithms being based on Chow and Liu's method for constructing dependence trees [5]). Actually, we could say that it is a combination of both algorithms, with some additional features.

As it can be noticed, this algorithm has three main different parts: in step 1, we compute the degrees of dependency between all pairs of nodes. The second part, step 2, represents the tree skeleton construction, and the last one, from step 3 until the end, performs the orientation of the edges in the tree, finally making up a polytree.

Several remarks have to be made about these three parts. First, the measure used to establish the dependency between nodes (which is, in some sense, analogous to the functions usually employed in IR systems for measuring the similarity between the terms in the

1. For every pair of nodes $\alpha, \beta \in U$, being $U$ the set of nodes, do
1.1. Compute $Dep(\alpha, \beta|\emptyset)$.
2. Build a maximum weight spanning tree $G$, where the weight of each edge $\alpha$–$\beta$ is

$$Dep(\alpha, \beta) = \begin{cases} Dep(\alpha, \beta|\emptyset) & \text{if } \neg I(\alpha, \beta|\emptyset) \\ 0 & \text{if } I(\alpha, \beta|\emptyset) \end{cases} \quad (1)$$

3. For every triplet of nodes $\alpha, \beta, \gamma \in U$ such that $\alpha$–$\gamma$, $\gamma$–$\beta \in G$ do
3.1. If $Dep(\alpha, \beta|\emptyset) < Dep(\alpha, \beta|\gamma)$ and $\neg I(\alpha, \beta|\gamma)$ then direct the subgraph $\alpha$–$\gamma$–$\beta$ as $\alpha \to \gamma \leftarrow \beta$.
4. Direct the remaining edges without introducing new head to head connections.
5. Return G.

Figure 1: Algorithm for Learning a Polytree

collection) is

$$Dep(\alpha, \beta|\emptyset) = \sum_{\alpha_i \beta_j} p(\alpha_i \beta_j) \ln\left(\frac{p(\alpha_i \beta_j)}{p(\alpha_i) p(\beta_j)}\right) \quad (2)$$

That is, the Kullback-Leibler cross entropy (also called Mutual information measure), which measures the dependency degree between two variables $\alpha$ and $\beta$ (which is equal to zero if $\alpha$ and $\beta$ are marginally independent, and such that the more dependent $\alpha$ and $\beta$ are, the greater $Dep(\alpha, \beta|\emptyset)$ is). The probabilities $p(\alpha_i, \beta_j)$ are estimated from the inverted file by counting frequencies. So, for example, $p(\alpha_1, \beta_0)$ is the probability that the term $\alpha$ appears in a document and not $\beta$, and $p(\alpha_1)$ is the probability of occurrence of term $\alpha$ (we are assuming that a term $\alpha$ is relevant for a given document if it appears in that document). Here, we use, as the RP algorithm does, the marginal cross entropy as dependency measure. The PA algorithm uses a combination of $Dep(\alpha, \beta|\emptyset)$ and the conditional dependency degrees (conditional mutual information measures) $Dep(\alpha, \beta|\gamma)$, for every other node $\gamma$:

$$Dep(\alpha, \beta|\gamma) = \sum_{\alpha_i \beta_j \gamma_k} p(\alpha_i \beta_j \gamma_k) \ln\left(\frac{p(\alpha_i \beta_j \gamma_k) p(\gamma_k)}{p(\alpha_i \gamma_k) p(\beta_j \gamma_k)}\right)$$

The reason why we have not combined the marginal dependency of two terms with the conditional dependencies of these two terms conditioned to the rest of terms, has been that, due to the great amount of terms in a collection, the computation of the conditional dependencies, although it has to be carried out only once, has been proved extremely time-consuming but also a big storage is needed. These two obstacles have leaded us to use only the marginal dependency as the dependency measure; methods to include some conditional dependencies in certain cases are being studied.

The next step is the tree skeleton construction. If we assume that the computed dependency values are link weights in a graph, this algorithm gets a maximum weight spanning tree (MWST), i.e. a tree where the sum of the weigths of its links is maximum. We considered two different methods to obtain the MWST: Kruskal's and Prim's algorithms [6], although, finally, we opted for the latter. The fundamental reason for this choice was that Kruskal's algorithm is suggested for sparse graphs and not for complete ones, as our case is, because it requires sorting all the computed dependencies. Prim's method begins with a one-node tree, and at each step adds to the tree the link between a node inside the tree and a node outside it having the maximum dependency degree, until $n - 1$ links have been adjoined, where $n$ is the number of nodes in the graph.

Due to the great number of terms that there are generally in a collection, the values of the dependencies are very low in general, and sometimes the algorithm does not have any good choice and selects as the highest value among all the dependencies being considered a very low value, adding the corresponding link to the tree. The problem lies in the fact that the two linked nodes are almost more independent than dependent, and therefore the model we are building loses accuracy with respect to the original one.

To solve this problem, the algorithm, once it has selected a new link $\alpha$–$\beta$ to be added to the tree, performs an independency test between $\alpha$ and $\beta$ (namely a Chi Square test with one degree of freedom based on the own value of $Dep(\alpha, \beta|\emptyset)$ [16]); then it really adds this link to the tree only if the independency test fails (in the algorithm in Figure 1, this is denoted as $\neg I(\alpha, \beta|\emptyset)$). This is equivalent to redefine the weights of the links, using eq. (1) instead of eq. (2). In this way, we can obtain a non-connected tree, i.e., a forest, as the result of this step.

Once the skeleton is built, the last part of the learning algorithm deals with the orientation of the tree, getting as a result a polytree. This process is based on the PA algorithm: in a head to head pattern $\alpha \to \gamma \leftarrow \beta$, the instantiation of the head to head node $\gamma$ should normally increase the degree of dependency between $\alpha$ and $\beta$, whereas in a non-head to head pattern such as $\alpha \leftarrow \gamma \to \beta$, the instantiation of the middle node $\gamma$ should produce the opposite effect, decreasing the degree of dependency between $\alpha$ and $\beta$. So, we compare the degree of dependency between $\alpha$ and $\beta$ after the instantiation of $\gamma$, $Dep(\alpha, \beta|\gamma)$, with the degree of dependency between $\alpha$ and $\beta$ before the instantiation of $\gamma$, $Dep(\alpha, \beta|\emptyset)$, and direct the edges toward $\gamma$ if the former is greater than the latter. Finally, the algorithm directs the remaining edges without intro-ducing new head to head connections. This strategy produced, in our preliminary experiments, structures where several nodes had a great number of parents; this fact leads to have very big probability tables and, as a consequence, it causes problems of storage and reliability (in the estimation of these tables). For that reason we have restricted a bit the rule that produces head to head connections, by including another condition in the antecedent: we want to be sure that if we decide to include a head to head connection $\alpha \to \gamma \leftarrow \beta$, then the nodes $\alpha$ and $\beta$ are not conditionally independent given $\gamma$. So, we also test this condition, once again using a Chi Square test of independency based of the value $Dep(\alpha, \beta|\gamma)$ (in this case with two degrees of freedom).

At last, and once the polytree structure has been built, the algorithm has to compute (from the inverted file) the prior probabilities for the root nodes, and the conditional probabilities of the rest of the nodes, given all their parents, and the Bayesian network is completely specified.

## 4 EXPERIMENTATION

In this section we are going to briefly describe the material used in our experiments (databases and software), as well as the experimental design and the obtained results.

### 4.1 STANDARD TEST COLLECTIONS AND SOFTWARE USED

Our experiments in query expansion using a Bayesian network-based thesaurus have been carried out with a set of three standard test collections. It is very common in IR that the experiments are made over standard test collections in order to compare several retrieval models. These collections are composed of a set of documents, a set of queries submitted by users and, finally, a set of relevance judgements, i.e. the set of documents which are relevant to each query. This last set is used to measure the retrieval effectiveness.

Three well-known standard collections have been the basis on which our experiments have been run: Adi, Cranfield and Medlars. Table 1 shows their basic characteristics. These collections were obtained from the Computer Science Department ftp site at Cornell University, as well as the Smart IR Software [27]. We have chosen these three collections because their characteristics are quite different. In particular, with respect to the number of terms, they cover a wide range of situations, going from a relatively small to a large number of terms.

Smart, as we said in section 2, is an IRS which implements the vector space model. An example of a typical representation of the $i^{th}$ document or query is $((t_{i1}, w_{i1}); (t_{i2}, w_{i2}); \ldots; (t_{in}, w_{in}))$, where each pair is a term and its weight.

| Collection | Adi | Cranfield | Medlars |
|---|---|---|---|
| Subject | Inform.Sci. | Aeronautics | Medicine |
| Documents | 82 | 1398 | 1033 |
| Terms | 828 | 3857 | 7170 |
| Queries | 35 | 225 | 30 |

Table 1: Characteristics of the Test Collections

Smart has different weighting schemes. The weighting process is composed of three phases:

1.- From indexing, the number of times that a term occurs in the document or in the query is associated to each term. This value is called *term frequency* $(tf)$. Also, the weight can be 0 if the term does not appear or 1 if it does. These weights can be normalised.

2.- Modification of the $tf$, normalised or not, using information obtained from the whole collection. The objective of this modification is to increase the weight of those terms rarest in the collection, and decrease the weight of the most common terms.

3.- Finally, a normalization process can be carried out over the entire vector.

Smart represents the three steps of the weighting process by three characters, each one representing a different weighting strategy in the whole weighting construction. In our experiments, we have used the $tf$ weight without normalisation, no modification of the $tf$ and no normalisation of the entire vector. In the Smart notation, our weighting scheme is 'nnn'. We have chosen it because is the 'default' option of Smart (although we know that it is not the best weighting scheme).

### 4.2 THE QUERY EXPANSION PROCESS. EXPERIMENT DESIGN

Given a query submitted to our system, the query expansion process starts placing the evidences in the learnt polytree. This action means looking for the terms that appear in the query in the polytree nodes and setting their states to '*the term is relevant*'. After that, a propagation process is carried out. As our network is a polytree, we can use an exact and efficient inference method to propagate the probabilities [19]. As a result of the propagation, the probability that a term is relevant, given that all the terms in the query are relevant, is obtained for each node. The next step is to get those terms with highest probability. To select the terms to be added to the query we use a threshold value that establishes a lower limit, i.e., those terms whose posterior probability is higher than the given threshold are chosen to be incorporated in the query, setting up the expanded query. The $tf$ weights of the added terms are precisely their posterior probabilities. Once the new query vector is completely created, Smart takes that new query and runs a retrieval, getting a set of retrieved documents.

Summing up, the whole expansion process is divided into two parts: the learning and the propagation phases. The former has as arguments the document collection (the learning file) and the confidence level for the two independency tests carried out. The latter has two arguments: the Bayesian network and the threshold to select the additional terms. With this set of arguments we have designed a battery of experiments for each collection, by using different confidence levels (90%, 95%, 97.5%, 99% and 99.5%) and different thresholds (0.5, 0.6, 0.7, 0.8 and 0.9)

So, for each collection a set of five thesauri has been created, each one with a different confidence level. The next step is to expand the set of standard queries of each collection using each one of the five thesauri of that collection. Each expansion is run five times, once per threshold, getting five new query files per thesaurus. Once twenty-five new sets of queries per collection are ready, Smart runs a retrieval and evaluates the retrieval effectiveness.

### 4.3 RESULTS

Tables 2, 3 and 4 display some of the results obtained for the three collections from all our experiments. In each table, precision/recall data are presented showing precision at ten standard recall levels and the average of precision at all ten levels, using Smart with and without query expansion (Q.E.); we also show the difference as the percent change from the baseline (Smart). As suggested in [28], a difference of 5% in average precision is generally considered significant and a 10% difference is considered very significant. The precision value displayed at each fixed recall level is the average for the given number of queries. The last two rows in each table represent recall and precision obtained from the retrieval of a fixed number of documents (we used the default value of Smart, which is 15).

For each collection, we show the results of using query expansion only for one experiment, as well as the average results over the twenty-five experiments. Nevertheless, for the Adi and Medlars collections we obtained better results than the baseline in all the twenty-five experiments, in terms of both the average precision values and number of documents retrieved. So, our query expansion method seems quite robust for these two collections. The best average precision values are more often achieved in low-medium thresholds (0.5-0.7). The precision values in our experiments for the ten recall points are always better than those ones in the original set of queries, establishing greater distances in medium-high recall values. In these two collections, there are not significant differences between the five networks, corresponding to the five confidence levels used in the independency tests.

| Recall level | Smart | Q.E.+ Smart | % change | Aver. 25 exper. |
|---|---|---|---|---|
| 0.1 | 0.4824 | 0.5052 | 4.73 | 0.4992 |
| 0.2 | 0.4343 | 0.4616 | 6.29 | 0.4642 |
| 0.3 | 0.4203 | 0.4365 | 3.85 | 0.4403 |
| 0.4 | 0.3640 | 0.3987 | 9.53 | 0.3916 |
| 0.5 | 0.3405 | 0.3878 | 13.89 | 0.3856 |
| 0.6 | 0.2775 | 0.3392 | 22.23 | 0.3053 |
| 0.7 | 0.2247 | 0.2761 | 22.87 | 0.2490 |
| 0.8 | 0.2143 | 0.2653 | 23.80 | 0.2423 |
| 0.9 | 0.1874 | 0.2417 | 28.98 | 0.2123 |
| 1.0 | 0.1863 | 0.2405 | 29.09 | 0.2111 |
| Average | 0.3132 | 0.3553 | 16.53 | 0.3401 |
| Recall | 0.5036 | 0.5983 | 18.80 | 0.5746 |
| Precision | 0.1524 | 0.1695 | 11.22 | 0.1679 |

Table 2: Results on the Adi Collection. Confidence Level: 95; Threshold: 0.5

| Recall level | Smart | Q.E.+ Smart | % change | Aver. 25 exper. |
|---|---|---|---|---|
| 0.1 | 0.5236 | 0.5150 | -1.64 | 0.4887 |
| 0.2 | 0.4158 | 0.4177 | 0.46 | 0.3926 |
| 0.3 | 0.3042 | 0.3105 | 2.07 | 0.2962 |
| 0.4 | 0.2452 | 0.2558 | 4.32 | 0.2415 |
| 0.5 | 0.2171 | 0.2289 | 5.44 | 0.2124 |
| 0.6 | 0.1687 | 0.1810 | 7.29 | 0.1675 |
| 0.7 | 0.1183 | 0.1263 | 6.76 | 0.1187 |
| 0.8 | 0.0931 | 0.1016 | 9.13 | 0.0927 |
| 0.9 | 0.0656 | 0.0696 | 6.10 | 0.0650 |
| 1.0 | 0.0573 | 0.0609 | 6.28 | 0.0569 |
| Average | 0.2209 | 0.2267 | 4.62 | 0.2132 |
| Recall | 0.3397 | 0.3450 | 1.56 | 0.3284 |
| Precision | 0.1591 | 0.1636 | 2.83 | 0.1576 |

Table 3: Results on the Cranfield Collection. Confidence Level: 97.5; Threshold: 0.9

However, for the Cranfield collection, it can be observed how the threshold used to select the terms to expand the queries is an important parameter in the retrieval performance: the greater the threshold is the better the results are. The expansions are not good at all for lower thresholds but they improve when this limit is higher: we only obtain better results than the baseline using high thresholds (0.8 and 0.9) (this explains the rather poor averages displayed in the last column in table 3). On the other hand, there are not remarkable differences between the results from the different networks, having a very similar behaviour. These results suggest that in this case query expansion is useful when the number of terms added to each query is rather small (this fact is probably due to the characteristics of this collection).

| Recall level | Smart | Q.E.+ Smart | % change | Aver. 25 exper. |
|---|---|---|---|---|
| 0.1 | 0.6389 | 0.7042 | 10.22 | 0.6788 |
| 0.2 | 0.5810 | 0.6110 | 5.16 | 0.6016 |
| 0.3 | 0.5204 | 0.5598 | 7.57 | 0.5505 |
| 0.4 | 0.4561 | 0.4938 | 8.27 | 0.4865 |
| 0.5 | 0.3725 | 0.3996 | 7.28 | 0.4013 |
| 0.6 | 0.2887 | 0.3381 | 17.11 | 0.3297 |
| 0.7 | 0.2403 | 0.2894 | 20.43 | 0.2786 |
| 0.8 | 0.1956 | 0.2367 | 21.01 | 0.2292 |
| 0.9 | 0.1534 | 0.1752 | 14.21 | 0.1725 |
| 1.0 | 0.0875 | 0.1115 | 27.43 | 0.1033 |
| Average | 0.3535 | 0.3919 | 13.87 | 0.3832 |
| Recall | 0.3161 | 0.3406 | 7.75 | 0.3356 |
| Precision | 0.4467 | 0.4800 | 7.45 | 0.4738 |

Table 4: Results on the Medlars Collection. Confidence Level: 97.5; Threshold: 0.7

So far we do not know exactly why our method performed for Adi and Medlars much better than for Cranfield. To solve this question will probably require an in-depth study of the characteristics of these collections (specificity or generality of queries and indexing terms, number of terms per query, etc).

In general, we may conclude that the improvement in retrieval effectiveness induced by our method is significant but moderate. This fact encourages us to continue studying this topic, in order to get more sensible improvements. Other conclusions are, on the one hand, that the confidence level for the independency test is not very relevant to the retrieval performance of the expanded queries, but, on the other hand, the threshold imposed to select those terms which are going to be added to the original query may be quite relevant. This suggests the necessity of developing some heuristics for automatically selecting the threshold according to the data in a given collection (for example taking into account the *inverse document frequencies* of terms in the queries). Anyway, it seems us that the improvement in retrieval performance is collection-dependent.

## 5  CONCLUDING REMARKS

We have developed a new query expansion method using a Bayesian network-based thesaurus (constructed automatically using a learning algorithm). Although we have applied it in combination with the well-known Smart system, our method could also be used as a preprocessing step for any other Information Retrieval System. The results obtained from the experiments show a moderate improvement of the retrieval effectiveness when using our query expansion technique. Anyway, we consider this work as a first step in a more general research centered on applying Bayesian networks to Information Retrieval.

Our short-term research lines are to develop new learning algorithms, completely adapted to the IR context, or modify existing ones and test them on the good testing ground of query expansion, helped by an IRS like Smart to run the retrievals. We will focus our long-term efforts on the development of an IRS completely based on Bayesian networks.

For our future developments, the main requirements that we are going to ask for are:

- We will look for more accuracy in the thesaurus learning algorithms, in the sense of creating more complicated structures than the polytree introduced in this paper, with the aim of being nearer to the real model defined by data, and at the same time taking into account the complexity of the problem. To carry out this idea, one of the first attempts will be the incorporation of conditional dependencies in the learning process.

- Incorporating documents to our models, creating structures which would represent the relationships between them, and learning the document network as precisely as is possible.

- Improving the performance of the propagation process, choosing the most adequate techniques and modifying them to reduce the time consumed in the propagation. For example, by making a preprocessing in which the probability tables were involved, so that at the moment of propagation the calculations would be minimum.

## Acknowledgments

This work has been supported by the Spanish Comisión Interministerial de Ciencia y Tecnología (CICYT) under Project n. TIC96-0781.